\documentclass[aps,prb,preprint,superscriptaddress]{revtex4-1}

\usepackage{graphicx}
\usepackage{bm}
\usepackage{amsmath}
\usepackage{amssymb}
\usepackage{hyperref}
\usepackage{color}

\begin{document}

\title{Thermoelectric transport properties of silicon: Towards an \emph{ab initio} approach}

\author{Zhao Wang}
\email{wzzhao@yahoo.fr}
\affiliation{LITEN, CEA-Grenoble, 17 rue des Martyrs, 38054 Grenoble Cedex 9, France}
\author{Shidong Wang}
\affiliation{Department of Mechanical Engineering and Materials Science, Duke University, Durham, NC 27708, USA}
\author{Sergey Obukhov}
\affiliation{Tomsk State Pedagogical University, 634061 Kievskaya 60, Tomsk, Russia}
\author{Nathalie Vast}
\affiliation{Ecole Polytechnique, Laboratoire des Solides Irradi\'es, CEA-DSM-IRAMIS, CNRS UMR 7642, 91120 Palaiseau, France}
\author{Jelena Sjakste}
\affiliation{Ecole Polytechnique, Laboratoire des Solides Irradi\'es, CEA-DSM-IRAMIS, CNRS UMR 7642, 91120 Palaiseau, France}
\author{Valery Tyuterev}
\affiliation{Tomsk State Pedagogical University, 634061 Kievskaya 60, Tomsk, Russia}
\author{Natalio Mingo}
\affiliation{LITEN, CEA-Grenoble, 17 rue des Martyrs, 38054 Grenoble Cedex 9, France}

\begin{abstract}
We have combined the Boltzmann transport equation with an {\it ab initio} approach to compute the thermoelectric coefficients of semiconductors. Electron-phonon, ionized impurity, and electron-plasmon scattering rates have been taken into account. The electronic band structure and average intervalley deformation potentials for the electron-phonon coupling are obtained from the density functional theory. The linearized Boltzmann equation has then been solved numerically beyond the relaxation time approximation. Our approach has been applied to crystalline silicon. We present results for the mobility, Seebeck coefficient, and electronic contribution to the thermal conductivity, as a function of the carrier concentration and temperature. The calculated coefficients are in good quantitative agreement with experimental results. 
\end{abstract}


\maketitle

\section{Introduction}
\label{section:S1}

Interest in thermoelectric materials has been rapidly growing in recent years. This is in part due to the new expectation of materials with a higher dimensionless figure of merit $ZT$ brought about by the nanotechnology revolution.\cite{Boukai:2008,Harman:2002,Hochbaum:2008,Venkatasubramanian:2001} From the theoretical point of view, it is important to be able to predict various thermoelectric properties without resorting to adjustable parameters. However, despite great advances in predicting the electronic structure of materials, calculation of thermoelectric transport properties \emph{from first principles} still presents a challenge, even in the case of simple bulk materials.

In some relevant works, attempts have been made to use a \emph{from first principles} description of the band structure, combined with the relaxation time approximation (RTA) of the scattering mechanisms.\cite{Wang:2009,Oh:2008,Huang:2008,Singh:2007,Wilson-Short:2007,Hazama:2006,Madsen:2006,Madsen:2006b,Ishii:2004,Thonhauser:2003} However, these approaches can still be considered as semi-empirical, since the scattering rates depend on adjustable parameters, and moreover rely on approximations for their energy dependence. A recent calculation of the mobility of SiGe alloys has been presented in Ref. \onlinecite{Murphy-Armando:2008}, where the scattering rates have also been computed \emph{from first principles} in the RTA. Regarding silicon, its mobility has been explored employing a parameter free approach \cite{Restrepo:2009} within the RTA\cite{Dziekan:2007,Yu:2008}, 
and its lattice thermal conductivity has been computed \emph{ab initio}, together with that of germanium and diamond.\cite{Broido:2007,Ward:2009}

Thus, previous \emph{ab initio} works largely relied on the RTA, and in most cases, the \emph{ab initio} aspect was limited to the calculation of electronic band structures. Our goal is to go beyond this stage by computing \emph{ab initio} scattering rates, and solving the BTE beyond the RTA. In the present study we are focussing on the thermoelectric properties of crystalline silicon, despite the fact that it is not a good candidate for practical thermoelectric applications. However, its electronic properties have been intensively investigated in the scientific literature, with a large amount of theoretical studies devoted to the calculation of the mobility of silicon using adjustable parameters.\cite{Jacoboni:1983,Fischetti:1991,Rode1972} Si constitutes an ideal system to test the ability of \emph{ab initio} methods to predict transport coefficients.

This paper is organized as follows. The general theory for the Boltzmann equation is presented in Section \ref{section:boltzmann}. The \emph{ab initio} method used to compute the band structure and the average deformation potentials for the intervalley electron-phonon scattering is described in Section \ref{section:abinitio}. Results are discussed in Section \ref{section:results}, and conclusions are drawn in Section \ref{section:conclusion}.

\section{Boltzmann transport equation}
\label{section:boltzmann}

The electron occupation distribution function $f$ has been obtained by solving the Boltzmann transport equation (BTE)\cite{Ashcroft:1976,Nag:1972} for steady states in presence of a uniform and static external electric field $\bm{E}$ and a temperature gradient $\nabla T$,\cite{Rode:1995}

\begin{eqnarray}
\label{eq:1}
- \frac{\partial f}{\partial \varepsilon} \bm{v} \cdot (e\bm{E}+\nabla T \frac{\varepsilon-\varphi}{T}) = \frac{-V} {8\pi^{3}} 
\int \left[ f(\bm{k})(1-f(\bm{k}^{'}))S(\bm{k},\bm{k}^{'})- f(\bm{k}^{'})(1-f(\bm{k}))S(\bm{k}^{'},\bm{k}) \right] d\bm{k}^{'}
\end{eqnarray}

\noindent where $\varepsilon$ is the band energy associated with a point $\bm{k}$ in the semiclassical phase space, $\varphi$ is the chemical potential, $\bm{v}$ is the electron velocity, $V$ is the unit cell volume, $S(\bm{k},\bm{k}^{'})$ is the probability of a transition from $\bm{k}$ to $\bm{k}^{'}$ per unit of time.

The linear approximation\cite{Nag:1972} consists in assuming that $f$ can be written as a linear function of the applied external fields. At low field, 

\begin{equation}
\label{eq:2}
f(\bm{k}) = f_{0}(\bm{k})+ e \bm{E} \cdot \bm{v}(\bm{k}) g(\bm{k}) + \frac{\nabla T}{T} \cdot \bm{v}(\bm{k}) g^{*}(\bm{k}) ,
\end{equation}

\noindent where $f_{0}$ is the equilibrium Fermi-Dirac distribution, $g(\bm{k})$ and $g^{*}(\bm{k})$ are the $1$st-order corrections to $f_{0}$ due to external electric 
field and temperature gradient. The approximation of non-degenerate statistics is not used in this work. The combination of Eq. \ref{eq:1} and Eq. \ref{eq:2} yields two linear matrix equations,

\begin{equation}
\label{eq:22}
\begin{array}{c}
\bm{\xi}=(\bm{S}^{out} - \bm{S}^{in}) \bm{g} , \\
\bm{\xi^*}=(\bm{S}^{out} - \bm{S}^{in}) \bm{g^*}.
\end{array}
\end{equation}

\noindent where $\xi=(8\pi^{3} v/V)(\partial f_{0} / \partial \varepsilon)$ and $\xi^{*}=(\varepsilon-\varphi)\xi$, and $\bm{\xi}$ and $\bm{g}$ are $n$-component vectors. $\bm{S}^{out}$ and $\bm{S}^{in}$ are $n\times n$ matrices where $n$ is the total number of $\bm{k}$ points. $\bm{S}^{out}$ is diagonal with $S^{out}_{i,i}= \sum_{j=1}^n \left[ S_{j,i}f_{0}^{j}+ S_{i,j}(1-f_{0}^{j}) \right]$. $\bm{S}^{in}$ reads $S^{in}_{i,j}=S_{i,j}f_{0}^{i}+ S_{j,i}(1-f_{0}^{i})$, with $S_{j,i}$ referring to $S(\bm{k}_{j},\bm{k}_{i})$.
An energy range of $0.13$ eV near the bottom of the conduction band has been discretized with 40 energy values. For each value of the energy $\varepsilon$, Eqs.~\ref{eq:22} have been solved for a set of $\bm{k}$ vectors on the surface of constant energy. In our calculation, discretization of the Brillouin zone is performed using the Gilat-Raubenheimer procedure (see next section). 

The matrix elements of $\bm{S}$ have been computed by considering different scattering mechanisms, including electronic interactions with intravalley and intervalley phonons, ionized impurities and plasmons, 
\begin{equation}
\label{eq:25}
S=S^{intra}+S^{inter}+S^{imp}+S^{plsm},
\end{equation}

\noindent where $S^{intra}$ represents the electronic scattering by a phonon of vanishing wave-vector $\bm{q}=\bm{0}$, for which, in the scattering process, the initial and final electronic states are located in the same valley. The term $S^{inter}$ represents the electronic scattering by a phonon of finite wave-vector $\bm{q} \ne \bm{0}$, the intervalley scattering, in which the initial and final electronic states belong to
two different valleys. In silicon, this mechanism involves a scattering from the valley $\Delta$, located near the $X$ point, to one of the $5$ other equivalent $\Delta^{'}$ valleys with $\bm{k}^{'}$=$\bm{k}+\bm{q}$. The matrix $S^{imp}$ represents the scattering probability of the electrons by 
the impurities. The latter three scattering processes $S^{intra}$, $S^{inter}$ and $S^{imp}$ have been calculated using models described in Jacoboni and Reggiani.
\cite{Jacoboni:1983} The deformation potentials used to compute $S^{inter}$ have been calculated \textit{ab initio} as described in the next section.
For electronic interactions with zone-center acoustic phonons $S^{intra}$, we have used the elastic approximation, assuming
the phonon energy to be negligible. The intervalley scattering is an inelastic process,
with scattering probability $S^{inter}$ which reads\cite{Jacoboni:1983}:
\begin{equation}
S^{inter}_{i,j}=\frac{\pi}{V\eta\omega_{ph}}\Phi^2N_{ph}\delta(\varepsilon_i-\varepsilon_j-\hbar\omega_{ph}).
\end{equation}
Here, $V$ is the unit cell volume, $\eta$ is the crystal mass density, $\omega_{ph}$ and $\Phi$ are the effective phonon frequency and deformation potential,
characterizing a particular intervalley scattering channel, $N_{ph}$ is phonon occupation number, $\varepsilon_i$ and $\varepsilon_j$ are the energies of the initial and final electronic states. In our calculations, the following approximation has been made:
\begin{equation}
f_0(\varepsilon_i)\delta(\varepsilon_i-\varepsilon_j\pm\hbar\omega_{ph})\approx \delta(\varepsilon_i-\varepsilon_j)f_0(\varepsilon_i\pm\hbar\omega_{ph})\frac{\rho(\varepsilon_i\pm \hbar\omega_{ph})}{\rho(\varepsilon_i)},
\end{equation}
where $\rho$ is the density of states.
Finally, $S^{plsm}$ stands for the electron-plasmon interaction probability, and has been calculated using Fischetti's approach.\cite{Fischetti:1991} 

Once the value of the distribution function $f$ was determined, the electrical conductivity $\sigma$ has been calculated as 

\begin{equation}
\label{eq:10}
\sigma = \frac{2e}{8\pi^{3}} \int g{(\bm{k})} \frac{{\bm{v} \cdot \bm{E}}}{\left| \bm{E} \right|} d \bm{k}.
\end{equation}

The electron mobility is given by 

\begin{equation}
\mu = \frac{\sigma}{eN},
\end{equation}

\noindent where $N$ is the carrier concentration.

The Seebeck coefficient (thermopower) $Q$ has been calculated from the solution of $g$, using

\begin{equation}
\label{eq:29}
\sigma Q = \frac{2e}{8\pi^{3}T} \int (\varepsilon-\varphi) g \bm{v} \cdot \frac{\bm{E}}{\left| \bm{E} \right|} d \bm{k}, 
\end{equation}

%

Finally, the expression used to evaluate the electronic contribution to the thermal conductivity $\kappa$ was 

\begin{equation}
\label{eq:14}
\kappa = - T \sigma Q^{2} + \frac{2e}{8\pi^{3}} \int (\varepsilon-\varphi) g^{*} \bm{v} \cdot \frac{\bm{E}}{\left| \bm{E} \right|} d \bm{k} .
\end{equation}

\section{Ab initio calculations}
\label{section:abinitio}

BTE is coupled with the band structure of silicon computed \emph{from first principles}, and to the electron-phonon coupling constants obtained \emph{ab initio} for the intervalley scattering.\cite{Sjakste:2007c,Sjakste:2006,Sjakste:2007,Sjakste:2008,Tyuterev:2010} The calculations have been performed respectively within the density functional theory\cite{Hohenberg:1964,Kohn:1965} (DFT) and the density functional perturbation theory\cite{Baroni:1987,Baroni:2001} (DFPT). The local density approximation (LDA) has been used for the exchange and correlation functional. We have used the pseudopotential and plane-wave approach to solve the Kohn-Sham equations. For silicon, the pseudopotential of Refs. \onlinecite{Giannozzi:1991,Tyuterev:2010} has been used, and the size of the plane-wave basis set has been limited with a cut-off energy of $45$ Ry. Finally, a $4 \times 4 \times 4$ Monkhorst-Pack grid has been used to sample the Brillouin zone (BZ), yielding $10$ non-equivalent $\bm{k}$ points in the irreducible BZ. We have obtained an equilibrium lattice parameter $a$ of $5.40$ \AA~for silicon, and the conduction band minimum $\Delta$ in the irreducible BZ was at $\mathbf{k}_s= (0,0,k_0)\frac{2\pi}{a}$, with $k_0=0.84$ in our calculations\cite{Tyuterev:2010} and $k_0=0.85$ in the scientific literature.\cite{Madelung:1982} 

Each isosurface has been defined by a given energy value above the conduction band edge $\varepsilon$, and has been determined \emph{from first principles}. To obtain the $\bm{k}$ points forming the isosurface, we have first divided the irreducible BZ into a set of small parallelepipeds, $l_x = \mathbf{b}_x/N_x$, $l_y = \mathbf{b}_y/N_y$, $l_z = \mathbf{b}_z/N_z$, where $\mathbf{b}_{\alpha}$ and $N_{\alpha}$ are the unit reciprocal lattice vectors and the number of segments in the direction ${\alpha}$, respectively. In each elemental parallelepiped, a representative k-point at the given value of $\varepsilon$ has been obtained with Brent's method.\cite{Brent:1972} The weight of the selected k-point has been computed as the area of the energy surface confined in the parallelepiped. This area has been calculated using the method proposed by Gilat and Raubenheimer.\cite{Gilat:1965} The constant energy surface in the parallelepiped has been approximated by a plane intersecting with the parallelepiped. The plane has been chosen according to the energy gradient at the representative k-point. The details of the method can be found in Ref.~\onlinecite{Gilat:1965}. When the parallelepiped was small enough, the plane has proved to be a very good approximation to the constant energy surface.

\begin{table}[thb]
\begin{center}
\begin{tabular}{llllllll}\hline\noalign{\smallskip}
\hline\noalign{\smallskip}
&      &\multicolumn{2}{c}{Jacoboni\footnotemark[1]} &    \multicolumn{4}{l}{This work } \\
   & phonon  & $D$   & $\omega$  & & $D$   & \multicolumn{2}{c}{$\omega$} \\
   & label   & (eV/\AA)& (K)     & & (eV/\AA) &  (K)  & (THz) \\
\hline\noalign{\smallskip}
$g$-processes &&&           & &      & &      \\
   & TA   & 0.5  & 140 & &0.61\footnotemark[2] & 138 & 2.88   \\
   & LA   & 0.8  & 215 & &1.22\footnotemark[2] & 226 & 4.72    \\
   & TO   &    &   & &1.26\footnotemark[2]     & 698 & 14.56 \\
   & LO   & 11.0 & 720 & &4.18\footnotemark[3] & 720 & 15.01  \\
\hline
$f$-processes &&&           & &      & &          \\
   & TA$_1$ & 0.3  & 220 & &0.18\footnotemark[2] & 217 & 4.52   \\
   & TA$_2$ &    &   & &0.20\footnotemark[2]     & 263 & 5.48   \\
   & LA   & 2.0  & 550 & &1.12\footnotemark[3]   & 521 &10.86  \\
   & LO   &    &   & &2.08\footnotemark[2]       & 569 &11.86  \\
   & TO$_1$ & 2.0  & 685 & &2.40\footnotemark[2] & 658 &13.72  \\
   & TO$_2$ &    &   & &4.24\footnotemark[3]     & 669 &13.95  \\
\end{tabular}
\footnotetext[1] {From Table VI of Ref.~\protect\onlinecite{Jacoboni:1983}.}
\footnotetext[2] {First-order transition (see text).}
\footnotetext[3] {0$^{th}$ order transition.}
\caption{Deformation potentials for the intervalley scattering: effective deformation potentials and phonon frequencies of Ref.~\protect\onlinecite{Jacoboni:1983}
(columns 3 and 4)
and \emph{ab initio} average deformation potentials (column 4, the value of $D$ has been averaged over states in the final valley for $f$-processes, and over both initial and final valleys for $g$-processes\protect\cite{Vast:Note:2010c})
and phonon frequencies (last two columns).
}
\label{fig:inter_scattering}
\end{center}

\end{table}

In BTE, the implementation of the electronic scattering by the short-wavelength phonons, $S^{inter}$, has been performed with average deformation potentials computed \emph{ab initio} (Table~\ref{fig:inter_scattering}). Details about the $f$-TA (transverse acoustical) and $g$-LA (longitudinal acoustical) processes have been reported elsewhere.\cite{Tyuterev:2010} The latter processes are forbidden by the symmetry selection rules to zero$^{th}$ order, when the phonon wavevector $\bm{q}$ is strictly equal to the vector connecting the bottoms of the two valleys $\Delta$ or $\Delta^{'}$. The scattering processes are however allowed for phonon vectors that connect points in the neighbourhood of $\Delta$ or $\Delta^{'}$ (first-order processes). 
The matrix elements for these first-order scattering processes have been computed on a grid with a step of $0.02\frac{2 \pi}{a}$, and have been averaged by fixing the initial electronic states at the $\Delta$ point and by summing the matrix elements over final electronic states close to the $\Delta^{'}$ point. The above-defined grid restricts the intervalley scattering to the processes involving energy values up to $100$ meV above the conduction band edge. Average deformation potentials of $0.2$ eV/\AA~ for $f$-TA and $1.2$ eV/\AA~for $g$-LA processes have been obtained.\cite{Tyuterev:2010} These values are close to the values of $0.3$ eV/\AA~ and $0.8$ eV/\AA~ used in Monte-Carlo simulations (column $3$ of Table~\ref{fig:inter_scattering}).\cite{Jacoboni:1983}

In contrast with the averaged values for the $f$-TA and $g$-LA processes however, our \emph{ab initio} averaged values for the intervalley scattering by other phonons can be widely different from the set of deformation potentials of Ref.~\onlinecite{Jacoboni:1983}. This appears to be especially true 
for the scattering by the $g$-LO phonon (Table~\ref{fig:inter_scattering}), and illustrates the need for parameter free calculations. Moreover, 
our calculations provide additional values for the scattering involving $g$-TO or $f$-LO phonons, which have been neglected so far in the modeling based on empirical deformation potentials (Table~\ref{fig:inter_scattering}). Last but not least, \emph{ab initio} calculations enable us to discriminate between the deformation potentials of the transverse phonons, which have been considered as degenerate in previous works.\cite{Jacoboni:1983} While this approximation is valid for the $f$-TA intervalley processes, because the \emph{ab initio} values of $f$-TA$_1$ and $f$-TA$_2$ are very similar, some caution must be taken with the scattering involving $f$-TO$_1$ or $f$-TO$_2$ phonons, for which we have found substantially different values of the deformation potentials (Table~\ref{fig:inter_scattering}). 

\section{Results and discussion}
\label{section:results}

\begin{figure}[ht]
\centerline{\includegraphics[width=13cm]{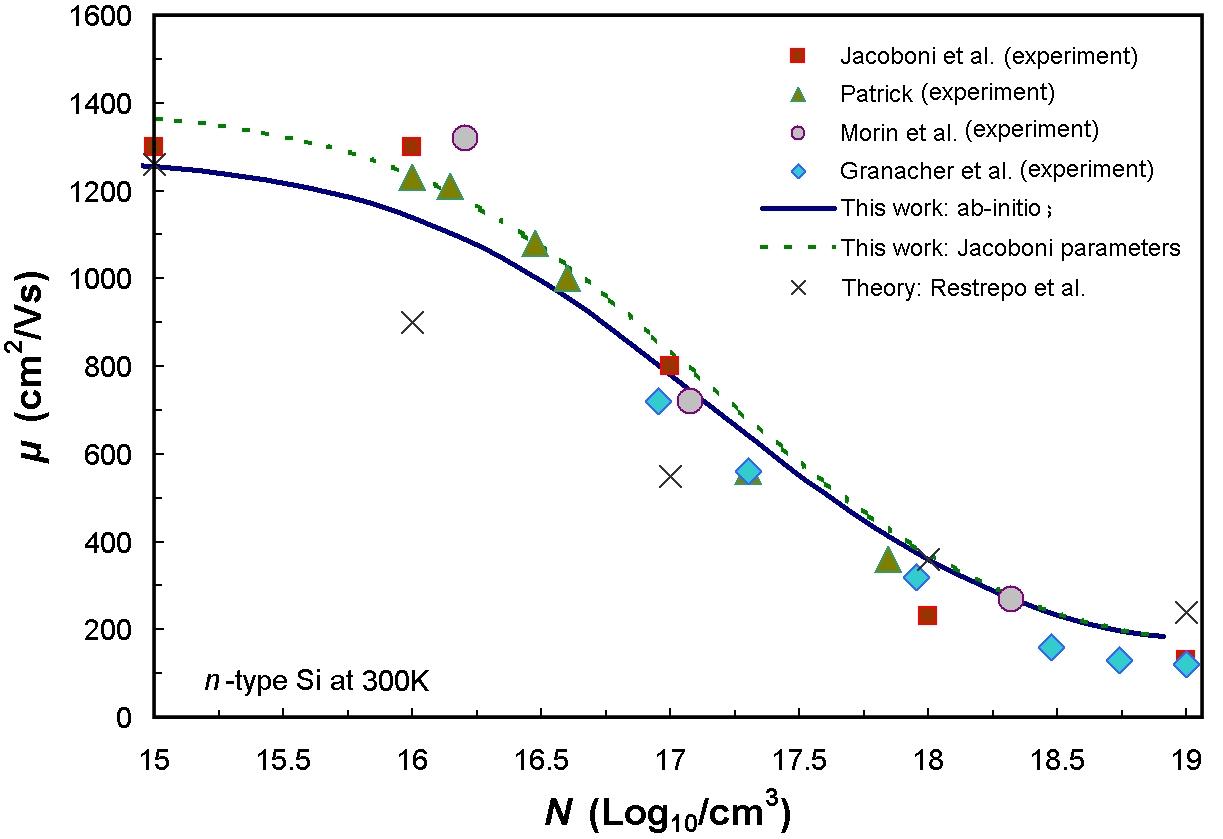}}
\caption{\label{fig:Mobilitycompare}
(Color-online) Electronic mobility $\mu$ \textit{versus} the carrier concentration $N$ for \textit{n}-doped Si at $T=300$~K. 
Solid and dashed lines are our results computed using deformation potentials for the intervalley scattering 
from our \textit{ab initio} calculations and from Ref. \onlinecite{Jacoboni:1983}, respectively. Crosses: theoretical results of Ref. \onlinecite{Restrepo:2009}.
The symbols stand for experimental data. Squares, triangles, circles and diamonds are results of Ref. \onlinecite{Jacoboni:1983}, \onlinecite{Patrick:1966}, 
\onlinecite{Morin:1954} and \onlinecite{Granacher:1967} respectively.}
\end{figure}

To make a valid comparison with the experimental results, the effects of all of the scattering mechanisms described in section \ref{section:boltzmann} have been taken into account. In Fig. \ref{fig:Mobilitycompare}, the electronic mobility has been computed with both sets of intervalley deformation potentials presented in Table \ref{fig:inter_scattering}, i.e. those computed \emph{ab initio} (solid line) and those of Jacoboni and Reggiani, Ref. \onlinecite{Jacoboni:1983} (dashed line). The parameters for the other scattering processes were taken from Ref. \onlinecite{Fischetti:1991}. In both cases we have used the DFT band structure. The comparison between our calculated electronic mobility and the experimental one (symbols) shows an excellent agreement over the whole range of carrier concentrations.

The mobility calculated using the empirical deformation potential values from Ref. \onlinecite{Fischetti:1991} is higher than that obtained from the DFT-calculated ones. In fact the empirical deformation potentials of Jacoboni and Reggiani, Ref. \onlinecite{Jacoboni:1983}, were adjusted to obtain the best fit to experimental data. 
We attribute the difference between the two curves to the increase in the number of scattering channels for the intervalley scattering computed \emph{ab initio} (see Table \ref{fig:inter_scattering}).  
In the high concentration region ($N>10^{18}$cm$^{-3}$), the electron-phonon interaction is negligible, and the two curves coincide. 

Comparing to the theoretical work of Ref. \onlinecite{Restrepo:2009} (Fig. \ref{fig:Mobilitycompare}, crosses), which used \emph{from first principles} calculations, our results give somewhat better agreement with the experimental data.
To our knowledge, the reason does not come from a difference in the treatment of the electron-phonon scattering, 
as the RTA approximation applied to equation (2) of Ref. \onlinecite{Restrepo:2009} amounts to the calculation of 
the average deformation potentials, as we have done in Table \ref{fig:inter_scattering}. 
Rather, several reasons can give rise to the observed differences. Firstly, the solution of the BTE was performed in the 
RTA approximation in the work \onlinecite{Restrepo:2009}, i.e. only $S^{out}$ matrix has been taken into account in eq. 
\ref{eq:22}. We have improved over the RTA approximation by taking into account $S^{out}-S^{in}$. Secondly, the elastic scattering by 
ionized impurities has been computed \emph{ab initio} in Ref. \onlinecite{Restrepo:2009}, whereas we still rely on the 
model of Jacoboni and Reggiani\cite{Jacoboni:1983}, adjusted by the best fit to the experimental data. Lastly, in contrast 
to work of Ref. \onlinecite{Restrepo:2009}, we have included electron-plasmon interaction as in Ref. \onlinecite{Fischetti:1991}, 
which becomes important in the high doping regime ($N>$10$^{17}$ cm$^{-3}$).  

\begin{figure}[bht]
\centerline{\includegraphics[width=13cm]{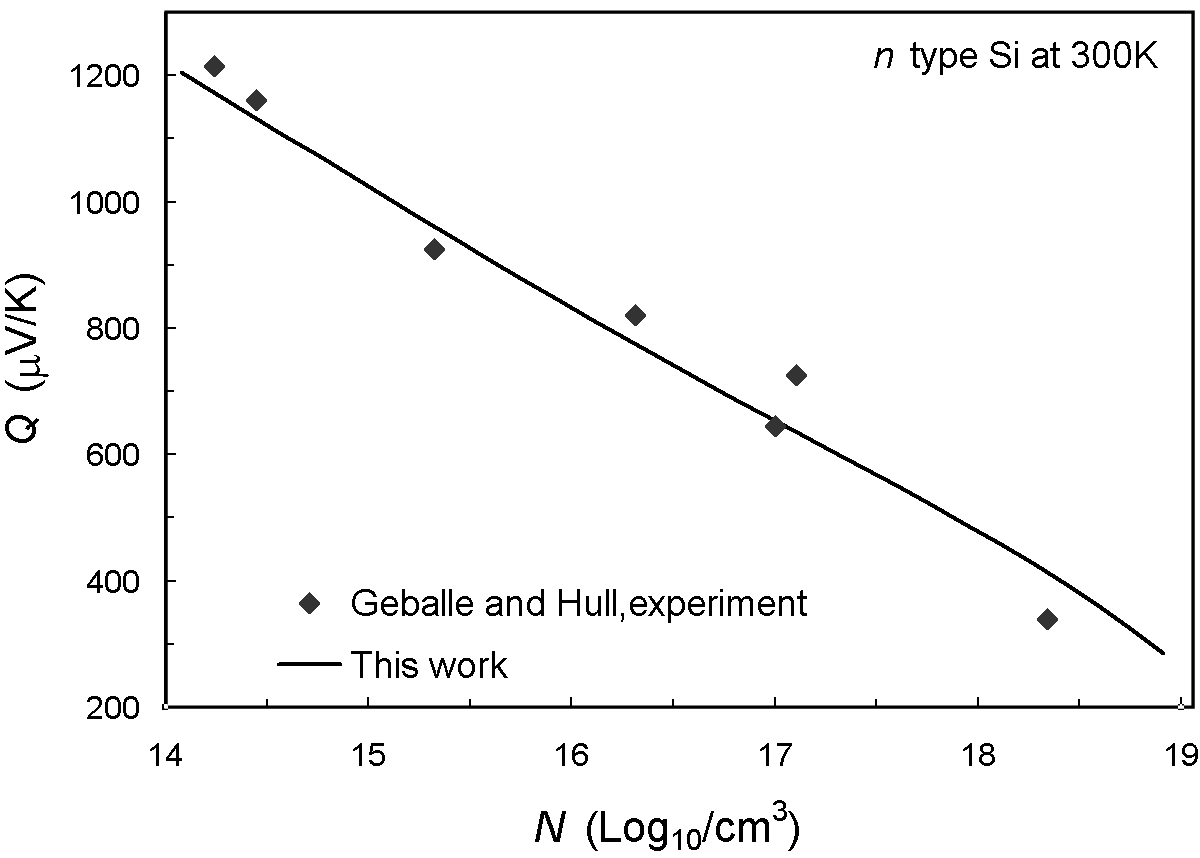}}
\caption{\label{fig:Seebeckcompare}
Seebeck coefficient $Q$ \textit{versus} carrier concentration $N$ for \textit{n}-doped Si at $T=300$~K. 
Solid line: this work. Diamonds: experimental results from Ref. \onlinecite{Geballe:1955}, where $N$ has been calculated by us as $N_{d}-N_{ac}$, where
$N_{d}$ and $N_{ac}$ are respectively the donor and acceptor concentrations.}
\end{figure}

Perfect agreement of our results with the existing experimental data\cite{Geballe:1955} is shown in Fig. \ref{fig:Seebeckcompare} for the Seebeck coefficient $Q$.
No significant difference in $Q$ has been found between the values calculated using the \textit{ab initio} deformation potentials and those obtained with the empirical parameters of Ref. \onlinecite{Jacoboni:1983}. 

\begin{figure}[thb]
\centerline{\includegraphics[width=16cm]{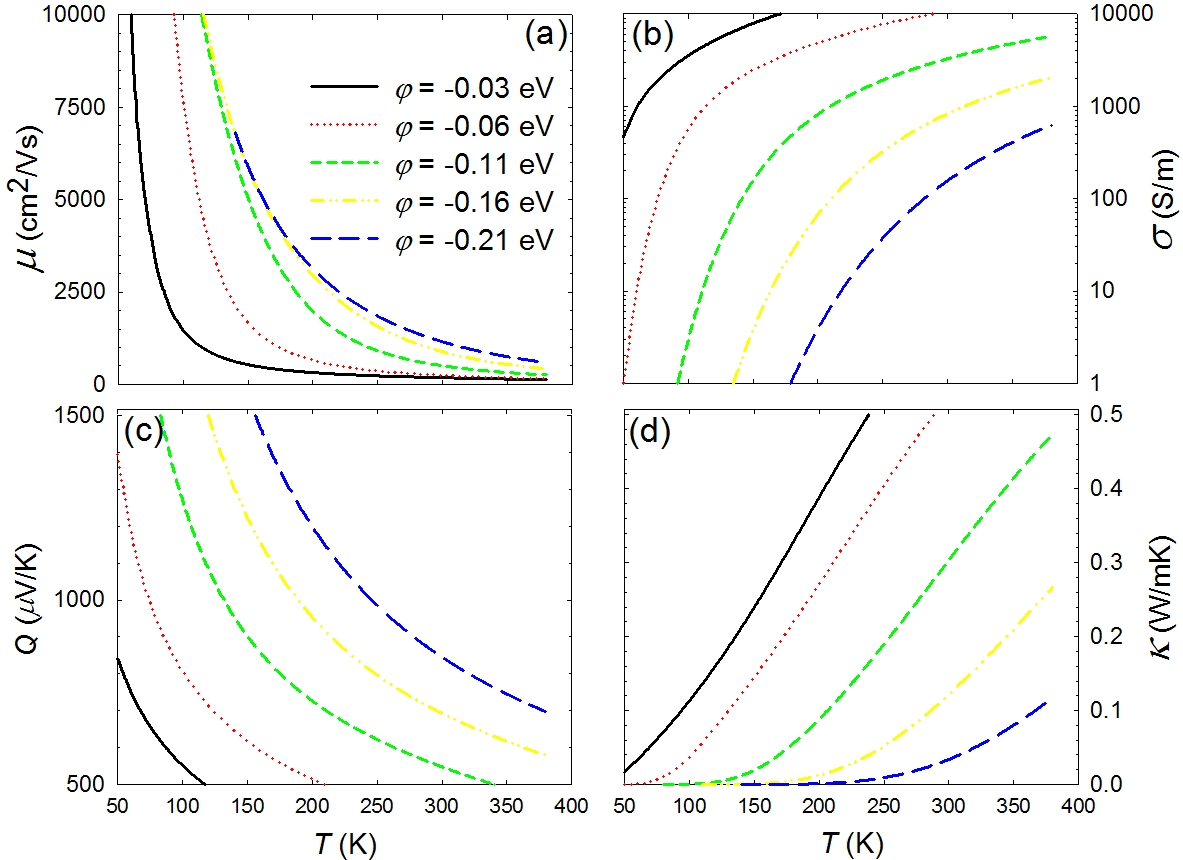}}
\caption{\label{fig:temperature}
(Color-online) Thermoelectric properties of \textit{n}-doped Si as a function of the temperature (T) for five values of the chemical potential $\varphi$. (a) electronic mobility; (b) electrical conductivity; (c): Seebeck coefficient (or thermal power); (d): electronic contribution to the thermal conductivity. 
For the calculation of $\varphi$ the energy of the bottom of the conduction band has been set to zero.}
\end{figure}

Thermal and electric transport properties of Si have been intensively studied in the past.\cite{Jacoboni:1983,Fischetti:1991} However, very few works gave thermoelectric coefficients over a wide range of doping concentrations and temperatures in a systematic way. In order to fill this gap, we have systematically computed electronic mobility $\mu$, electrical conductivity $\sigma$, Seebeck coefficient $Q$ and thermal conductivity $\kappa$ as a function of the temperature $T$ for different doping concentrations, which correspond to different values of the chemical potential $\varphi$. In Fig.\ref{fig:temperature} (a), significant increase of $\mu$ can be observed when the temperature decreases below $150$ K, in agreement with the inverse power scaling law $\mu \propto 1/T^{a}$, where $a$ is a positive number increasing with $\left| \varphi \right|$.\cite{Ziman:1960} The
electrical conductivity $\sigma=eN\mu$ increases with $T$, since more carriers are generated at higher temperatures for a given $\varphi$, despite the decrease of $\mu$ with temperature. It can be seen that, for a given $T$, $\sigma$ has the highest value when the value of $\varphi$ is close to the bottom of the lowest conduction band (panel b). The value of $Q$, the thermal power, decreases when the temperature increases at a given chemical potential (panel c), as expected from eq. \ref{eq:29}. The thermal conductivity $\kappa$ increases with the temperature at a given $\varphi$. This trend is however different from that observed for nanowires, where thermal transport is dominated by phonons.\cite{zhao2010} 

\begin{figure}[thb]
\centerline{\includegraphics[width=16cm]{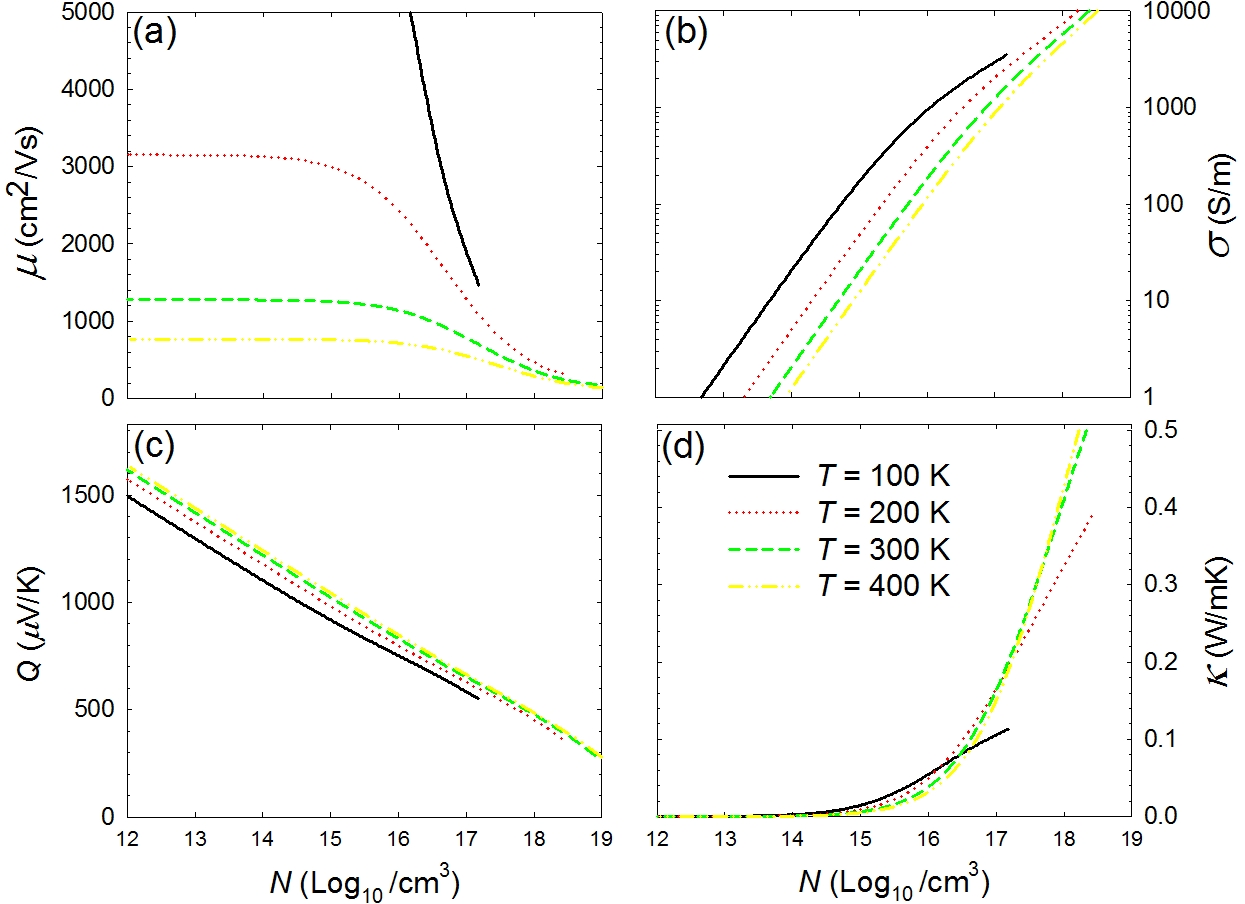}}
\caption{\label{fig:CN}
(Color-online) Thermoelectric properties of \textit{n}-doped Si \textit{versus} carrier concentration $N$ at four different values of the temperature. Panel a: electronic mobility; Panel b: electrical conductivity (on a logarithmic scale); Panel c: Seebeck coefficient or thermal power. Panel d: electronic contribution to the thermal conductivity.}
\end{figure}

We have also studied the effect of doping on the thermoelectric properties at given temperatures. The value of the electronic mobility $\mu$ is almost constant in the low concentration region of $N<10^{16}$cm$^{-3}$, and then it rapidly decreases when $N$ becomes larger than $10^{16}$cm$^{-3}$ (panel (a) of Fig.\ref{fig:CN}). It can be seen that $\sigma$ linearly increases with increasing $N$ for weak doping, since $\mu$ is constant in this doping range (panel (b) of Fig.\ref{fig:CN}). At higher carrier concentrations, the increase becomes less significant due to the decrease in the mobility. Furthermore, $Q$ decreases almost linearly with increasing $log{(N)}$ (panel c of Fig.\ref{fig:CN}). The electronic contribution to the thermal conductivity $\kappa$ also sensitively depends on the carrier density in the conduction band (panel (d) of Fig.\ref{fig:CN}). We can see that $\kappa$ is very weak in the weak doping domain ($N<10^{15}$cm$^{-3}$), as significant thermal transport by electrons can be observed only at high doping levels.

\section{Conclusion}
\label{section:conclusion}

In conclusion, we have combined the Boltzmann transport equation with 
\emph{from first principles} calculations of the electronic band structure and of the electron-phonon coupling constants for the intervalley scattering in silicon. This approach is developed to improve the calculation accuracy beyond the relaxation time approximation for the determination of thermoelectric coefficients of semiconductors. The computed electronic mobility and the Seebeck coefficient at room temperature have been compared to experimental data. Good quantitative agreement has been obtained. Furthermore, temperature and doping effects on thermoelectric coefficients have been investigated in a systematic way, and provide predictive data. 
\section*{Acknowledgments}
Results have been obtained with the (modified) Quantum Espresso package.\cite{Giannozzi:2009,Baroni:2001,Pwscf} We thank Paola Gava for the careful reading of the manuscript and many suggestions for its improvement. Useful discussions with M. Calandra, M. Lazzeri and F. Mauri are acknowledged. The authors acknowledge support from the ANR (project PNANO ACCATTONE), and computer time granted by GENCI (project 2210).
\end{document}